\documentclass[conference]{IEEEtran}
\usepackage[nocompress]{cite}
\usepackage[pdftex]{graphicx}
\graphicspath{{figures/}}
\usepackage{array}
\usepackage{listings}
\usepackage[frozencache=true,cachedir=minted-cache,newfloat=true]{minted}
\usepackage{xspace}
\usepackage{amsmath}
\usepackage{comment}
\usepackage{soul}
\usepackage[acronym]{glossaries}
\usepackage{glossaries-extra}
\glsdisablehyper %
\setabbreviationstyle[acronym]{long-short}

\usepackage{xurl}
\usepackage{hyperref}
\hypersetup{breaklinks=true,colorlinks,citecolor=blue,linkcolor=blue,urlcolor=blue}
\usepackage{cleveref}
\usepackage{subcaption}
\usepackage{booktabs}
\usepackage{pifont}
\usepackage{threeparttable}
\usepackage{circledsteps}
\usepackage{todonotes}
\usepackage{paralist}

\newif\ifAnon
\newif\ifWithCode

\begin{document}
\title{CacheSquash:\\Making caches speculation-aware}
\ifAnon
\else
\author{\IEEEauthorblockN{Hossam ElAtali}
\IEEEauthorblockA{University of Waterloo\\
hossam.elatali@uwaterloo.ca}
\and
\IEEEauthorblockN{N. Asokan}
\IEEEauthorblockA{University of Waterloo\\
asokan@acm.org}}
\fi

\pagestyle{plain}

\renewcommand{\paragraph}{\noindent\textbf}

\newcommand{\circletter}[1]{\raisebox{.6pt}{\textcircled{\raisebox{-.8pt} {#1}}}\xspace}

\newcommand{\feature}{CacheSquash\xspace}

\newcommand{\GeomeanIpcSpecSingleCore}{$-1.84\%$\xspace}
\newcommand{\GeomeanIpcSpecFourCore}{$0.90\%$\xspace}
\newcommand{\GeomeanIpcSpecAllCores}{$-0.48\%$\xspace}

\newcommand{\GeomeanIpcOverheadSpecAllCores}{$0.48\%$\xspace}

\newcommand{\GeomeanTicksSpeedupParsecMedSingleCore}{$-0.18\%$\xspace}
\newcommand{\GeomeanTicksSpeedupParsecMedFourCore}{$4.36\%$\xspace}
\newcommand{\GeomeanTicksSpeedupParsecMedAllCores}{$2.06\%$\xspace}

\newcommand{\GeomeanTicksOverheadParsecLargeSingleCore}{$0.20\%$\xspace}
\newcommand{\GeomeanTicksOverheadParsecLargeFourCore}{$0.54\%$\xspace}
\newcommand{\GeomeanTicksOverheadParsecLargeAllCores}{$0.37\%$\xspace}

\newcommand{\GstdSpecAllCores}{$1.10$\xspace}
\newcommand{\GstdParsecMedAllCores}{$1.07$\xspace}
\newcommand{\GstdParsecLargeAllCores}{$1.008$\xspace}

\newacronym{MSHR}{MSHR}{miss status holding register}
\newacronym{LLC}{LLC}{last-level cache}
\newacronym{CPU}{CPU}{central processing unit}
\newacronym[longplural=proofs-of-concept]{poc}{PoC}{proof-of-concept}
\newacronym{ISA}{ISA}{instruction set architecture}
\newacronym{PHT}{PHT}{pattern history table}
\newacronym{BCB}{BCB}{bounds check bypass}
\newacronym{ROB}{ROB}{reorder buffer}
\newacronym{ROP}{ROP}{return-oriented programming}
\newacronym[longplural=regions-of-interest]{roi}{ROI}{region-of-interest}
\newacronym{TLB}{TLB}{translation lookaside buffer}
\newacronym{RSB}{RSB}{return stack buffer}
\newacronym{STT}{STT}{Speculative Taint Tracking}
\newacronym{IPC}{IPC}{instructions-per-cycle}

\maketitle
\begin{abstract}
Speculation is key to achieving high CPU performance, yet it enables risks like Spectre attacks which remain a significant challenge to mitigate without incurring substantial performance overheads. These attacks typically unfold in three stages: access, transmit, and receive. Typically, they exploit a cache timing side channel during the transmit and receive phases: speculatively accessing sensitive data (access), altering cache state (transmit), and then utilizing a cache timing attack (e.g., Flush+Reload) to extract the secret (receive). 
Our key observation is that Spectre attacks only require the transmit instruction to execute and dispatch a request to the cache hierarchy. It \emph{need not complete} before a misprediction is detected (and mis-speculated instructions squashed) because responses from memory that arrive at the cache after squashing still alter cache state.

We propose a novel mitigation, \feature{}, that \emph{cancels} mis-speculated memory accesses. Immediately upon squashing, a cancellation is sent to the cache hierarchy, propagating downstream and preventing any changes to caches that have not yet received a response. This minimizes cache state changes, thereby reducing the likelihood of Spectre attacks succeeding. We implement \feature{} on gem5 and show that it thwarts practical Spectre attacks, with \emph{near-zero performance overheads}.

\end{abstract}

\section{Introduction}

Speculation is a fundamental technique employed in modern \glspl{CPU} to optimize performance by predicting and executing instructions ahead of time. Correct predictions eliminate stalls in the processor pipeline, providing significant performance gains. An incorrect prediction, or \emph{mis-speculation}, causes the offending instructions to be \emph{squashed}, their results discarded and the pipeline flushed to restart execution at the correct location.

While speculation is tightly integrated into \gls{CPU} cores, the cache hierarchy in modern \glspl{CPU} is still not \emph{speculation-aware}. This means that loads executed speculatively will always be processed to completion by the cache hierarchy, even if the load is found to be a mis-speculation and squashed before the processing is complete. Cache state changes caused by the processing persist even after mis-speculation is detected. This results in a low attack barrier for Spectre attacks~\cite{Kocher2018spectre}.

Spectre attacks exploit speculative execution to leak sensitive information, such as cryptographic keys. They train the \gls{CPU}'s branch prediction mechanism and then use it to transiently access architecturally-inaccessible secrets in memory. The attack consists of three steps: accessing the secret, transmitting it through a side channel (e.g., changing cache state), and receiving/extracting it from the side channel (e.g., probing cache state). Notably, the transmit step, which involves sending a request to the cache hierarchy, does \emph{not need to complete within the speculation window} for the attack to succeed. 
The core executes loads and issues the read requests. The cache hierarchy receives requests, and processes them to completion (potentially modifying cache state), irrespective of if/when mis-speculation is detected. Any changes to cache state are persistent; attackers can probe them to leak information.

Despite efforts to mitigate it, Spectre remains a relevant threat, challenging the balance between security and performance~\cite{kasper, inspectreGadget}. Existing solutions~\cite{invisiSpec,cleanupSpec} aim for strong security guarantees, but require significant performance (and area) overheads. In systems where performance is critical and attacker capabilities limited, such as network equipment~\cite{ciscoSpectre,juniperSpectre}, the tradeoff is unacceptable, and a lightweight hardening mechanism with little to no performance impact is preferred.

In this paper, we propose \feature, a technique to make the cache hierarchy speculation-aware by \emph{sending a cancellation to the cache hierarchy as soon as an outstanding load is squashed}. The cancellation propagates down the hierarchy, preventing cache state changes that have not yet occurred, thereby significantly reducing the likelihood of a successful Spectre attack. Unlike prior work, we make a trade-off in favor of performance. \feature hardens protection against Spectre-style attacks at \emph{near-zero overheads}. Furthermore, \feature only requires minimal control-logic changes to the \gls{CPU}'s load-store unit and the caches' miss-handling circuitry, with \emph{no additional state storage}, unlike prior filtering or ``cache-undo'' approaches~\cite{muontrap, cleanupSpec}. It is applicable to any \gls{ISA} and requires \emph{no changes to software or external hardware interfaces}.

We implement \feature in gem5 and evaluate its performance under different configurations on the SPEC CPU 2017 and PARSEC benchmarks. We also evaluate its efficacy against various Spectre \glspl{poc} and provide an analysis on its efficacy against real-world attacks.

Our contributions are:
\begin{enumerate}
\itemsep 0em
    \item \feature, an \textbf{\gls{ISA}-agnostic} mechanism for cancelling read requests upon squashing, requiring \textbf{no changes to software or external hardware interfaces} (\Cref{sec:design}),
    \item its implementation in gem5 (\Cref{sec:impl})%
    \footnote{We will open-source the implementation.
    },
    \item performance evaluations showing a negligible geometric mean overhead (\GeomeanIpcOverheadSpecAllCores) on SPEC CPU 2017, and a geometric mean \textbf{speedup} of \GeomeanTicksSpeedupParsecMedAllCores and overhead of \GeomeanTicksOverheadParsecLargeAllCores on the medium and large PARSEC benchmarks, respectively, (\Cref{sec:perf}), and
    \item case studies showing that \feature is \textbf{effective} against several Spectre \glspl{poc} (\Cref{sec:sec-eval}).
\end{enumerate}

\section{Background}

\paragraph{CPU caches.}
Caches are small, high-speed memory structures situated closer to the \gls{CPU} cores compared to main memory. They are designed to store frequently-accessed data and instructions, thus reducing the time the \gls{CPU} needs to fetch this information from the slower main memory.

Data is stored in caches in chunks called ``lines'' or ``blocks'', usually 64 bytes in size. Each cache line stores the data itself as well as a tag that identifies which address the data belongs to. When a cache receives a read or write request, it searches its tags for one matching the request. If a match is found, this is called a ``cache hit'', and the cache can respond with the data. Otherwise, a ``cache miss'' has occurred and must be handled by a \gls{MSHR}. \Glspl{MSHR} are in charge of keeping track of outstanding misses, issuing requests downstream, and servicing the misses once a response is received. Each active \gls{MSHR} is in charge of a single tag (i.e., 64-byte aligned address). Multiple misses with the same tag are added as ``targets'' to the same \gls{MSHR}. Concretely, if there is already an outstanding \gls{MSHR} with the same tag, the miss is added to it as an additional target. Once a response is received for this \gls{MSHR}, all its targets are serviced. If there is no matching outstanding \gls{MSHR}, an empty \gls{MSHR} is allocated to the miss. Caches have a fixed number of \glspl{MSHR} and a maximum number of targets per MSHR; if there are no empty \glspl{MSHR} to handle a new miss or the matching \gls{MSHR} has reached its maximum number of targets, the cache must stall.

\paragraph{Speculation.}
Modern processors employ speculative execution to improve performance by predicting and executing instructions ahead of time. This allows the processor to continue processing instructions even when there is a branch instruction whose outcome is uncertain. The \emph{speculation window} refers to the period during which instructions are executed speculatively. The larger the speculation window, the more instructions that can be executed before a squash occurs.

\paragraph{Cache timing attacks.}
Cache timing attacks, such as Flush+Reload~\cite{flushReload} and Prime+Probe~\cite{primeProbe}, exploit variations in the time it takes for a \gls{CPU} to access cached vs. uncached data. A cache hit takes less time to complete than a cache miss. An attacker can compare the time it takes to access a certain address %
to determine whether the data at this address was cached. If a process uses secret-dependent memory addresses, this can leak information about the secret to the attacker.

\paragraph{Spectre.}
Spectre~\cite{Kocher2018spectre} attacks are a class of side-channel attacks that exploit speculative execution. They use speculative loads to leak sensitive information across security boundaries. By manipulating the \gls{CPU}'s branch prediction mechanism, an attacker can force the execution of speculatively loaded instructions that access sensitive data. Even though these instructions are eventually discarded, they can leave traces in the cache that can be exploited to infer the sensitive data. In other words, speculatively executed memory instructions can cause \emph{persistent changes to cache state}.

\section{Problem description}
\subsection{Goals \& Objectives}\label{sec:goals}

\newcommand{\goal}[2]{%
    \vspace{0.5em}
    \par\noindent%
    \newdimen\protocolstepwidth%
    \protocolstepwidth=\linewidth%
    \addtolength\protocolstepwidth{-2\fboxsep}%
    \fbox{\begin{minipage}[c]{\protocolstepwidth}\textbf{#1}: \emph{#2}\end{minipage}}%
    \vspace{0.5em}%
}

Ideally, desiderata for speculation-aware caches are:

\goal{\hypertarget{req.PR}{R1}---Performance}{
     no negative run-time performance impact on realistic workloads.
}
\newcommand{\perfreq}{\hyperlink{req.PR}{R1}}

\vspace{-0.5em}

\goal{\hypertarget{req.SCR}{R2}---Software Compatibility}{
    require \emph{no changes} to software and be fully compatible with existing program binaries.
}
\vspace{-0.5em}
\goal{\hypertarget{req.HCR}{R3}---Hardware Compatibility}{
    require \emph{no changes} to external interfaces (e.g., DRAM) or hardware components, other than the \gls{CPU}.
}
\newcommand{\hcreq}{\hyperlink{req.HCR}{R3}}
\vspace{-0.5em}
\goal{\hypertarget{req.ICR}{R4}---ISA Compatibility}{
    be applicable to any \gls{ISA}.
}
\newcommand{\icreq}{\hyperlink{req.ICR}{R4}}
\vspace{-0.5em}
\goal{\hypertarget{req.ER}{R5}---Effectiveness}{
     reduce leakage of secret data through cache state changes.
}
\newcommand{\ereq}{\hyperlink{req.ER}{R5}}

We define a \emph{cache change} metric $CC$ %
for effectiveness (\ereq): %

\begin{equation}\label{eq:cache-change}
    CC = \frac{\sum_{i}^{K} N_i \times (K - i + 1) }{N_{total} \times \sum_{i}^{K} i}
\end{equation}

$K$ is the number of cache levels in the system. $N_{total}$ is the total number of squashed access and transmit instructions in a program. %
$N_{i}$ is the number of squashed access and transmit instructions that cause a change in the $i^{th}$ cache. 
We assign more weight to changes to caches closer to the \gls{CPU} because they are easier to exploit via cache timing~\cite{llcAttacksPractical}. 
$CC \in$ [0,1]. Any non-zero value implies that Spectre attacks may succeed.

\subsection{Threat model}
We consider the strongest Spectre threat model, where the attacker and the victim execute within the same process, sharing the same address space and having the same process context. The attacker is unable to directly access the victim's secret (e.g., due to in-process isolation mechanisms such as sandboxing), but can train the branch predictor to access the secret speculatively. This corresponds to ``same-address in-place'' (SA-IP) as defined by Canella et al~\cite{transientfail}. A defense that works under this strong threat model, is also secure under weaker threat models such as where the attacker and victim are in different processes.
We only consider cache-timing channels; other channels, e.g., contention-based channels, are out of scope.

\section{\feature{}: Design}\label{sec:design}

\begin{figure*}[t]
    \centering
    \begin{subfigure}[t]{0.33\linewidth}
        \centering
        \includegraphics[height=6.1cm,trim={0 1cm 0 0},clip]{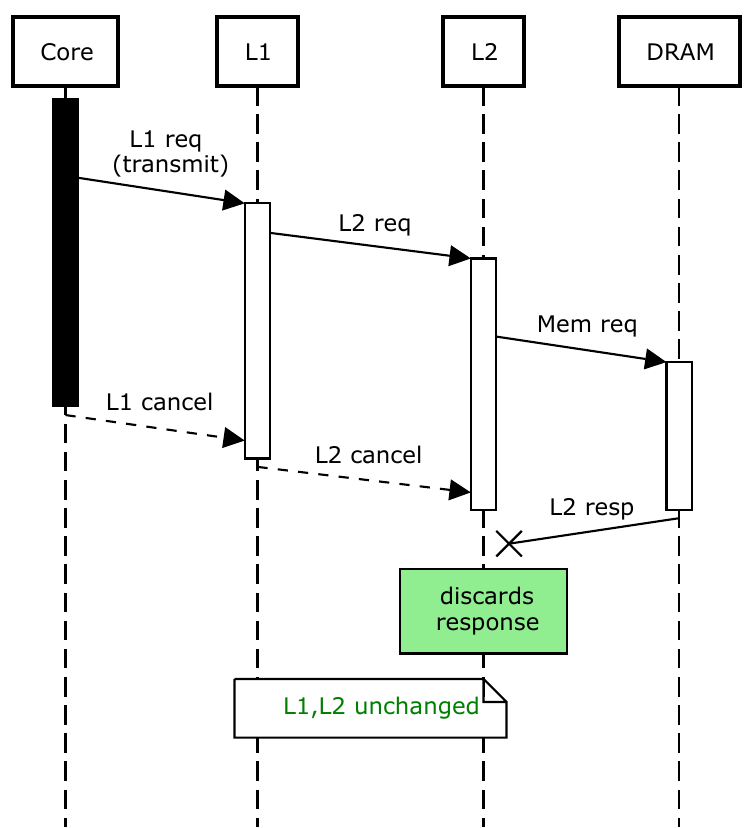}
        \caption{Best-case scenario}
        \label{fig:seq-best}
    \end{subfigure}%
    ~
    \begin{subfigure}[t]{0.33\linewidth}
        \centering
        \includegraphics[height=6.1cm,trim={0 1cm 0 0},clip]{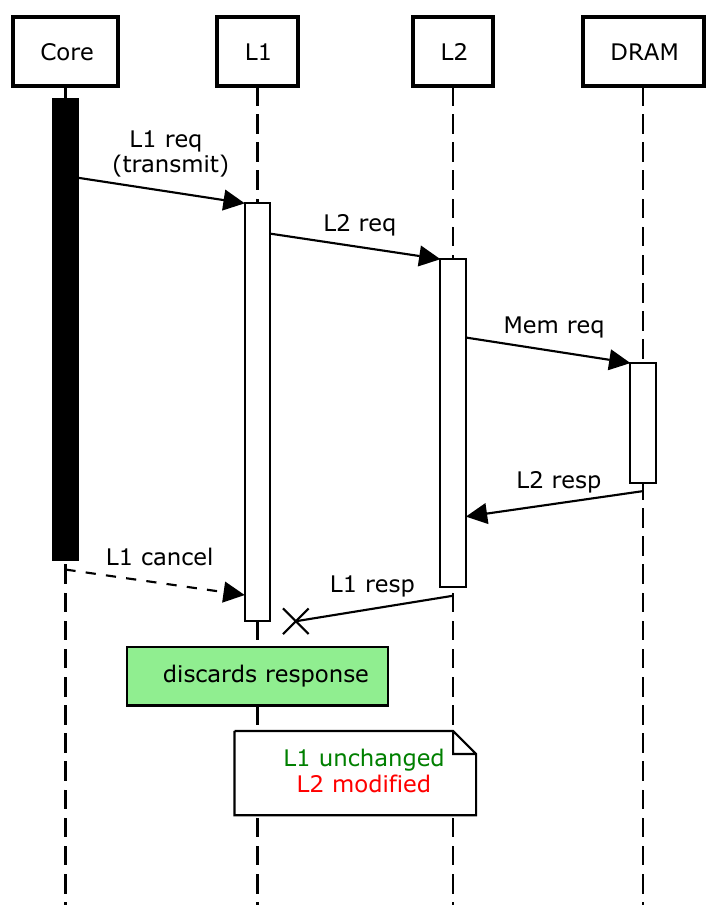}
        \caption{Intermediate}
        \label{fig:seq-mid}
    \end{subfigure}%
    ~
    \begin{subfigure}[t]{0.33\linewidth}
        \centering
        \includegraphics[height=6.1cm,trim={0 1cm 0 0},clip]{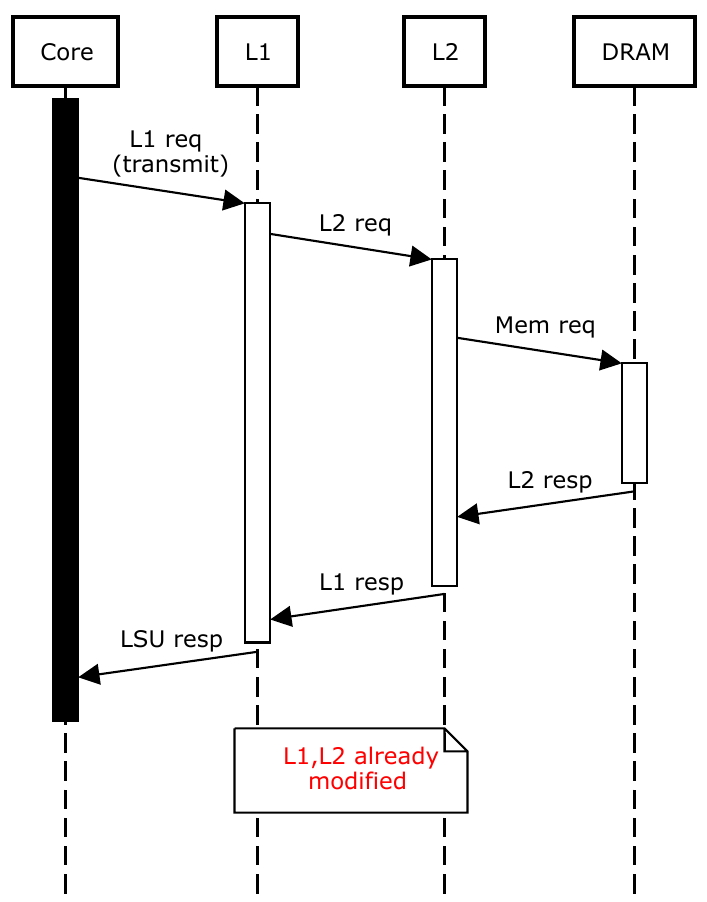}
        \caption{Worst-case}
        \label{fig:seq-worst}
    \end{subfigure}
    \caption{The solid black activation bar (for core) represents the speculation window. The hollow activation bars (for caches) represent lifetimes of the requests' \glspl{MSHR}. The hollow activation bar (for DRAM) represents the memory-only access latency.}
\end{figure*}

The idea behind \feature{} is to minimize cache state changes by issuing cancellations to outstanding speculative read requests as soon as they are squashed. Whenever a cache receives a cancellation for an outstanding request, it drops the request from its \glspl{MSHR} (and ignores any responses it receives for it in the future), and, if appropriate, forwards the cancellation to caches downstream. %

The final effect of the cancellation depends on the state of the read request/response within the cache hierarchy at the time the cancellation is sent. We present all possible cases in \Cref{fig:seq-best,fig:seq-mid,fig:seq-worst}. In the best case, \Cref{fig:seq-best}, the cancellation reaches the \gls{LLC}, L2, before it receives a response from memory. This prevents any changes to the cache hierarchy as any response received by the \gls{LLC} from memory is ignored; subsequently, the \gls{LLC} does not provide further responses to caches upstream (which have already cancelled the request and the corresponding evictions\footnote{Evictions do not require special handling as they are only completed when the response arrives, rather than when the \gls{MSHR} is first allocated.} themselves).

In the worst case, \Cref{fig:seq-worst}, the cancellation is either never made (because the response is received by the \gls{CPU} core before the speculation window ends), or it reaches L1 after it has received a response. If the \gls{CPU} has more than one cache level, other intermediate cases between \Cref{fig:seq-best} and \Cref{fig:seq-worst} can occur: \Cref{fig:seq-mid} shows %
a cancellation reaching L1, but not L2, before the response; only L2 is modified by the request.
If this request is from a Spectre transmit instruction, 
only attacks targeting the \gls{LLC} can succeed; those targeting L1 will not.

\subsection{Cache flow chart}
\begin{figure}[t]
    \centering
    \includegraphics[width=0.8\linewidth,trim={0.7cm 0.7cm 0.7cm 1cm},clip]{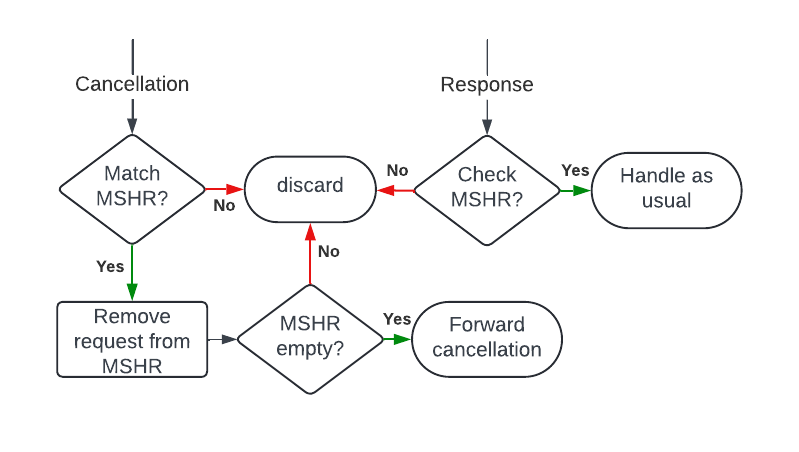}
    \caption{%
    \feature flow chart.}
    \label{fig:cache-flow}
\end{figure}

\Cref{fig:cache-flow} shows the \feature{} flowchart. It works with any cache coherence protocol%
. Beside the addition of handling cancellations, which is only relevant when a cache has an outstanding request to the level below it, the rest of the coherence protocol remains unchanged. %
Outstanding requests can either be reads (if the block is currently in the invalid state) or upgrades (if the block is valid, but does not have the required permissions, e.g., writable). As soon as we receive a cancellation, we only need to check whether there is a corresponding \gls{MSHR} (MatchMSHR), and if so, remove the cancelled request from it. If the MSHR then becomes empty (i.e., no other requests are waiting for this cache block), we can send a cancellation to the lower cache level. For a response, if a matching \gls{MSHR} is not found, it is discarded (CheckMSHR).

\subsubsection{MatchMSHR}%
MatchMSHR searches the \gls{MSHR} queue for a match, i.e., one that has the same cache block address. The circuitry to perform this search already exists in modern caches. It is required to check if incoming misses match an outstanding \gls{MSHR}, and coalesce requests to the same block. It is possible for cancellations to find no matching \gls{MSHR} due to simultaneous transfers of responses and cancellations.

\emph{Simultaneous response and cancellation.} %
Cache buses can have several channels, allowing simultaneous bidirectional communication, e.g., TileLink~\cite{tilelink} or Arm AMBA CHI~\cite{amba-chi}. This means that a cancellation can be received while a response is being sent. This can occur in two cases: 1) if the original request hits in this cache, or 2) the original request misses, but, before the cancellation arrives, the cache receives a response and the \gls{MSHR} is serviced and freed. The cancellation in both cases will have no corresponding \gls{MSHR} and therefore must be discarded. This requires an \gls{MSHR} search to detect.

\subsubsection{CheckMSHR}%
Without \feature{}, \glspl{MSHR} are locked to a single downstream request until a response is received. The downstream request contains an \gls{MSHR} index which is copied into the response by the downstream cache. This makes it easy to determine the exact \gls{MSHR} corresponding to a response. With \feature{}, this assumption no longer holds. The cache must double-check the designated \gls{MSHR} when receiving a response to ensure that it has not been freed and possibly reassigned to another block. This can happen either due to simultaneous response and cancellation transfer (as described above), or due to responses from memory.

\emph{Responses from memory.}
\feature{} does not require modifications to DRAM or memory controllers (satisfying \hcreq). Cancellations are not supported on the memory bus; \glspl{LLC} must therefore not forward cancellations to memory. Thus, a request from the \gls{LLC} to memory will eventually receive a response, even if it is cancelled in all caches. \glspl{LLC} need to detect whether the \gls{MSHR} in the response still corresponds to the original request, hence the need for CheckMSHR.

\subsection{Forwarding cancellations upstream}\label{sec:forwardcancel}
Cache coherence protocols ensure that requests for a cache line get an up-to-date response. The protocol can probe upstream caches for dirty cache lines, causing them to allocate an \gls{MSHR} while they probe caches further upstream. With \feature, forwarding cancellations is thus required to ensure these \glspl{MSHR} are freed.
Cache coherence protocols are mainly split into two groups: snooping-based and directory-based.
In snooping, caches ``snoop'' the bus to detect requests that require their intervention. This is usually done using additional circuitry and can span multiple cache levels. For \feature{}, this circuitry can also be used to snoop on cancellations.
In directory-based protocols, a directory tracks the caches holding each block. A cache receiving a cancellations must consult its directory and forward a cancellation to all caches with an outstanding \gls{MSHR} of a corresponding target.

\subsection{TLBs \& I-Caches}
An important cache-like structure in \glspl{CPU} is the \gls{TLB}, which caches virtual-to-physical address translations. 
Prior work has also shown that \glspl{TLB} can be used to leak information in a way similar to cache timing attacks~\cite{TLBleed}. Therefore, with \feature{}, we also send cancellations for mis-speculated address translations that result in page-table walks, reducing changes to \gls{TLB} state.%

The instruction cache is also affected by (mis-)speculation since the branch predictor speculates on which path of instructions will be executed. To minimize instruction cache pollution, we also send cancellations to the instruction cache when an instruction fetch is mis-speculated.

\section{\feature{}: Implementation}\label{sec:impl}

We implement \feature{} on gem5~\cite{gem5}, a cycle-accurate computer system simulator. gem5 includes a speculative out-of-order \gls{CPU} model, O3, and supports arbitrary cache configurations.
It provides two separate cache hierarchy implementations: classic and ruby. Ruby caches are newer and allow configurable cache coherence protocols. However, they currently do not support cache maintenance operations, such as flushing, and therefore cannot be used with the Flush+Reload cache timing attack. As this is the attack used by the majority of the publicly available Spectre \glspl{poc}, we choose to implement \feature{} on the classic caches.
The classic caches have a fixed snooping-based cache coherence protocol, so we implement cancellation snooping as described in \Cref{sec:forwardcancel}.

\paragraph{MatchMSHR \& CheckMSHR.}
The logic for MatchMSHR already exists for handling cache misses. We use the same latency of searching the \gls{MSHR} queue for cancellations. CheckMSHR adds new functionality. However, since no search is required, only a simple check, we assume it is combinational logic and can be performed in the same cycle. CheckMSHR thus incurs no additional latency.

\paragraph{O3 CPU.}
In addition to modifying the caches, we add \feature{} support to gem5's O3 model. Whenever an instruction is squashed, the load-store unit checks if there are any outstanding memory requests for the instruction, and if so, sends a cancellation to the cache hierarchy. Note that we do not make any changes to \gls{ISA}-specific \gls{CPU} models; \feature{} is \gls{ISA}-agnostic, satisfying \icreq.

\section{Security evaluation}\label{sec:sec-eval}

\subsection{Case studies}

We present case studies with two Spectre variants to empirically show the effectiveness of \feature{}. We analyze the first (\Cref{sec:safeside:pht}) to show the \gls{CPU} events occuring throughout a Spectre attack and how \feature{} affects them. For the second, we only report our findings for brevity; the attacks use the same access-transmit-receive mechanism. Note that since cancellations can be used for any squashed memory request, regardless of the speculation condition, \feature{} is equally applicable to \emph{all Spectre variants}.

All \glspl{poc} define a secret string as the target of the attack. For each character of the string, all \glspl{poc} continue attempting to leak the character until the extracted value matches certain criteria, e.g., value is a valid English ASCII character. 
This means that when attacks are successful, the program terminates quickly. On the other hand, if attacks do not succeed, the program runs until it times out. The default timeout periods for the \glspl{poc} are infeasibly long when run on gem5. We therefore shorten all timeouts to allow the simulation to complete in a reasonable amount of time. To ensure a fair comparison, we verify that, without \feature{}, both \glspl{poc} are still able to leak the secret using the shortened timeouts.

\subsubsection{Google SafeSide -- Spectre PHT}\label{sec:safeside:pht}
SafeSide~\cite{safeside} is a Google code repository containing several Spectre and Meltdown~\cite{Lipp2018meltdown} \glspl{poc}. We use the \texttt{spectre\_v1\_pht\_sa} \gls{poc}, which mistrains the \gls{PHT} and then exploits it to achieve a \gls{BCB}. The \gls{PHT} is a component of the \gls{CPU}'s branch predictor in charge of guessing whether a branch will be taken. The \gls{poc} uses Flush+Reload to transmit and receive the secret.

\renewcommand\fcolorbox[4][]{\textcolor{olive}{\strut#4}}
\begin{listing}[ht]
\begin{minted}[
xleftmargin=7pt,
frame=lines,
framesep=1mm,
baselinestretch=0.9,
fontsize=\footnotesize,
numbersep=5pt,
linenos
]
{gas}
# bounds check
404bec: jae    404bbf <main+0xca>
404bee: movsbq 0x0(%r13,%rax,1),%rax
404bf4: imul   $0x71,%rax,%rax
404bf8: add    $0x64,%rax
# access secret: data[local_offset]
404bfc: movzbl %al,%eax
404bff: add    $0x1,%rax
404c03: mov    %rax,%rsi
404c06: shl    $0x6,%rsi
404c0a: add    %rsi,%rax
404c0d: shl    $0x6,%rax
404c11: add    0x28(%rsp),%rax
# transmit: timing_array[secret]
404c16: movzbl (%rax),%eax
404c19: jmp    404bbf <main+0xca>
\end{minted}
\caption{\texttt{spectre\_v1\_pht\_sa} x86 assembly extract.}\label{lst:safeside-pht-asm}
\end{listing}

\Cref{lst:safeside-pht-asm} shows the x86 disassembly of the speculatively executed instructions. Lines 3-15 are executed speculatively until the failed bounds check on line 2 is detected and the instructions are squashed. Before the squash occurs, the secret is accessed (line 7) and transmitted across a side channel by modifying the cache state (line 15). Extracting the secret from the cache state is done non-speculatively afterwards.
We now describe four experiments to evaluate the \gls{poc} with and without \feature{}.

\begin{figure}[t]
    \centering
    \hspace{1cm}\includegraphics[width=0.6\linewidth]{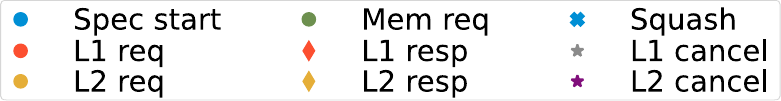}
    \includegraphics[width=0.9\linewidth,trim={0 0.2cm 0 2.47cm},clip]{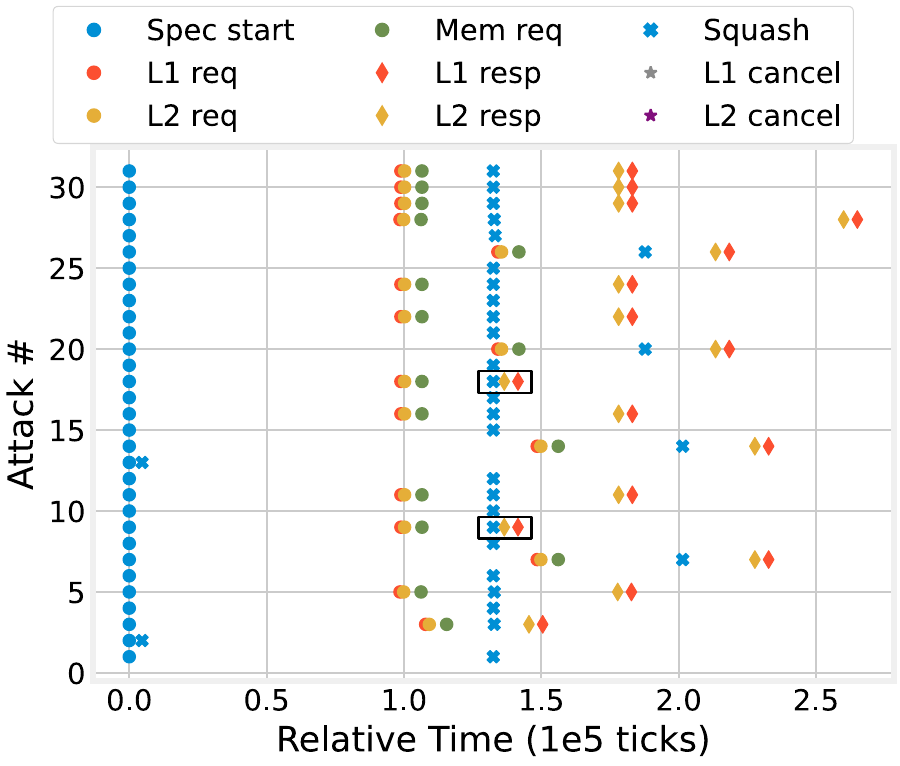}
    \caption{\gls{CPU} events for \textbf{transmit} instructions of all Spectre attack instances from \textbf{Experiment 1}. Each row represents a single attack, the y-axis showing the attack number. All event times are relative to speculation start: for each attack, events are shifted to make speculation start at t=0.}
    \label{fig:safeside-pht-plot-noCMR}
\end{figure}

\begin{table}[ht]
    \centering %
    \begin{tabular}{@{}lrr@{}}
    \toprule
        \textbf{Parameter} & \textbf{C1} & \textbf{C2}  \\
        \midrule
        Core count                  & 2      & 2        \\
        Core frequency (GHz)        & 3      & \color{red}{0.1}         \\
        Private L1I/D size (kB)     & 32     & 32       \\
        Shared L2 size (kB)         & 512    & 512      \\
        L1I/L1D/L2 associativity    & 8/8/16 & 8/8/16   \\
        L1I/L1D/L2 latency (cycles) & 4/4/14 & \color{red}{80/80/80}    \\ \bottomrule
    \end{tabular}
    \caption{Configurations used in case study experiments. For C1, realistic values are used for L1 and L2 based on Intel IceLake~\cite{agnerMicro}. C2 is modified from C1 to intentionally make \feature{} ineffective. For main memory, we use 3GB of dual channel DDR4-2400.}
    \label{tab:gem5-configs}
\end{table}
\paragraph{Experiment 1 -- C1, Baseline.} %
We first run the \gls{poc} in gem5 using the O3 model and the C1 gem5 configuration shown in \Cref{tab:gem5-configs} without \feature{}. The \gls{poc} is able to extract the entire secret. We dump all load-store unit and cache events from gem5, and identify and extract events related to pairs of the access and transmit instructions in \Cref{lst:safeside-pht-asm}. Each pair represents a single Spectre attack. \Cref{fig:safeside-pht-plot-noCMR} plots relevant events for the transmit instruction of each attack, showing that for all attacks where the transmit instruction is executed, the squash occurs before any response is returned to the \gls{LLC}. Further, in almost all of those cases, there is a significant delay between the squash occurring, and the \gls{LLC} receiving a response. This provides an opportunity for cancellations, and hints that \feature{} might prevent this attack (see our next experiment). Exceptions to this are the cases highlighted with black boxes. 
These can pose a problem because cancellations might not reach  L2 in time before the response from memory. We do not find this in our following experiments with C1, but we force this situation to occur in Experiment 4%
.

In attack 1 ($y=1$ in \Cref{fig:safeside-pht-plot-noCMR}), the transmit instruction is never executed. This occurs because the secret is not yet cached and the access instruction does not complete in time to allow the transmit instruction to execute. But subsequent attacks (e.g., attack 3) can access the now-cached secret quickly and thus have enough time to execute the transmit. In attack 3, the access instruction misses in L1  but hits in L2.

\begin{figure}[t]
    \centering
    \hspace{0.8cm}\includegraphics[width=0.6\linewidth]{figures/legend.pdf}
    \includegraphics[width=0.9\linewidth,trim={0 0.2cm 0 2.61cm},clip]{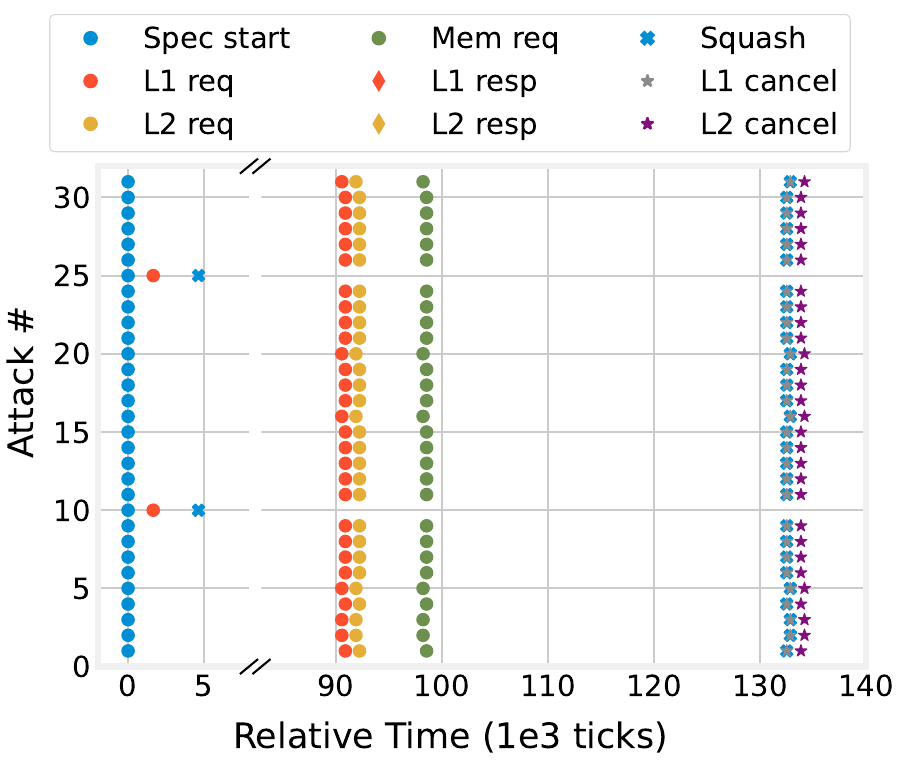}
    \caption{\gls{CPU} events for \textbf{access} instructions of Spectre attacks \#1--31 from \textbf{Experiment 2}. Attacks \#32-102 are not shown. Note the \textbf{break} in the x-axis. Cancellations have the same latency as regular requests. Typically, cancellations reach both L1 and L2 caches before a response is received. The secret is thus almost never cached in a Spectre attack.}
    \label{fig:safeside-pht-plot-CMR-access}
\end{figure}

\paragraph{Experiment 2 -- C1, \feature{}.} 
We run the \gls{poc} with C1 and \feature{} enabled, and indeed find that the program times out without leaking any secrets. Our analysis leads to an interesting discovery: \emph{no transmit instructions are executed}. Upon further investigation, we find that this occurs because \feature{} cancels the \emph{access instructions}, preventing even the first step of the attack. In its entire run, the \gls{poc} only finds the secret in the L1D cache twice, and in both cases, speculation ends before the transmit is executed. In all other cases, the attack has the same result as attack 1 in Experiment 1 (\Cref{fig:safeside-pht-plot-noCMR}). This is shown in \Cref{fig:safeside-pht-plot-CMR-access}, where we plot the access instructions instead, and see that they hit in L1D only twice. Note that because Experiment 2 times out, the total number of attacks (102) is much larger than in Experiment 1 (31). However, for clarity, we only plot attacks \#1--31.

\begin{figure}[t]
    \centering
    \hspace{0.8cm}\includegraphics[width=0.6\linewidth]{figures/legend.pdf}
    \includegraphics[width=0.9\linewidth,trim={0 0.2cm 0 2.61cm},clip]{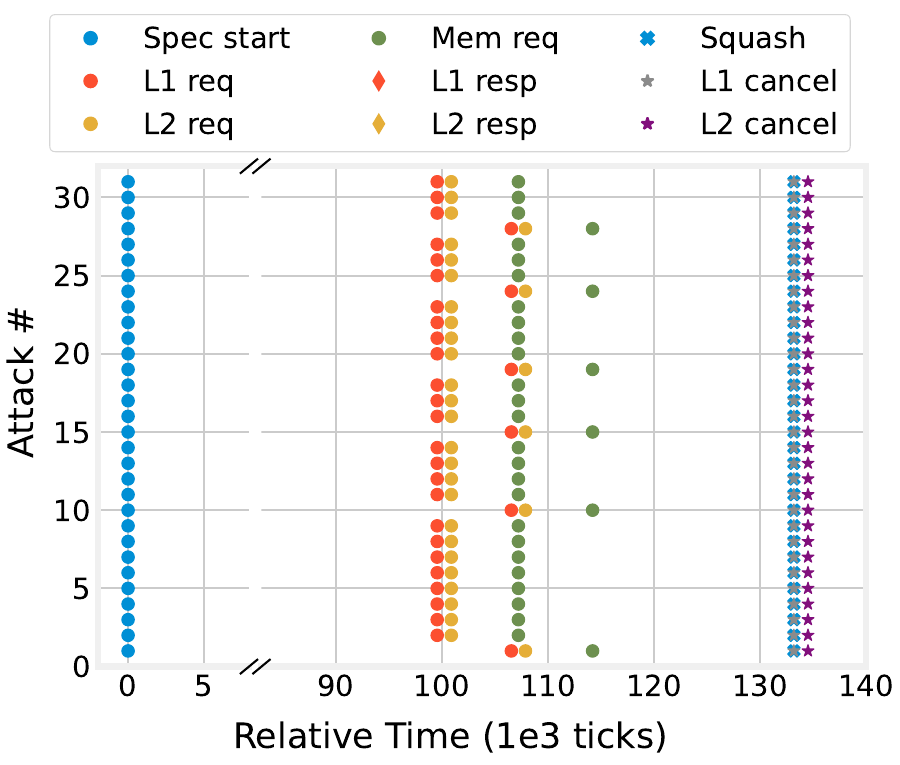}
    \caption{\gls{CPU} events for \textbf{transmit} instructions of attacks \#1--31 from \textbf{Experiment 3}. In all cases, the cancellations reach L1 and L2 caches before a response is received. The transmit instruction thus never succeeds in changing cache state.}
    \label{fig:safeside-pht-plot-CMR-cachedSecret}
\end{figure}

\paragraph{Experiment 3 -- C1, \feature{}, secret cached.} %
Here, we want to test the effectiveness of \feature{} even when the secret is cached and the access instruction completes. We modify the \gls{poc} to access the secret non-speculatively on every iteration. Running the \gls{poc} again results in a timeout with no secrets leaked. Our analysis confirms that the access instructions now hit in L1D, and the transmit instructions execute. We show all relevant events for the transmit instructions in \Cref{fig:safeside-pht-plot-CMR-cachedSecret}.
Despite the execution of the transmit, no Spectre attack succeeds. This is because the cancellations always reach the caches before the response, even for the last-level L2 cache, hence preventing any changes to cache state.

\paragraph{Experiment 4 -- C2, \feature{}, secret cached.} %
Here, we intentionally change the system configuration to C2 to reduce the effectiveness of \feature{}. We drastically increase the L1 and L2 latencies, and reduce the \gls{CPU} clock frequency to 100MHz, shown in red in \Cref{tab:gem5-configs}. Reducing the clock frequency effectively increases the speculation window %
relative to main memory latency (as main memory uses a separate clock), making responses from memory more likely to arrive within the speculation window.

\begin{figure}[t]
    \centering
    \hspace{0.8cm}\includegraphics[width=0.6\linewidth]{figures/legend.pdf}
    \includegraphics[width=0.85\linewidth,trim={0 0.2cm 0 2.61cm},clip]{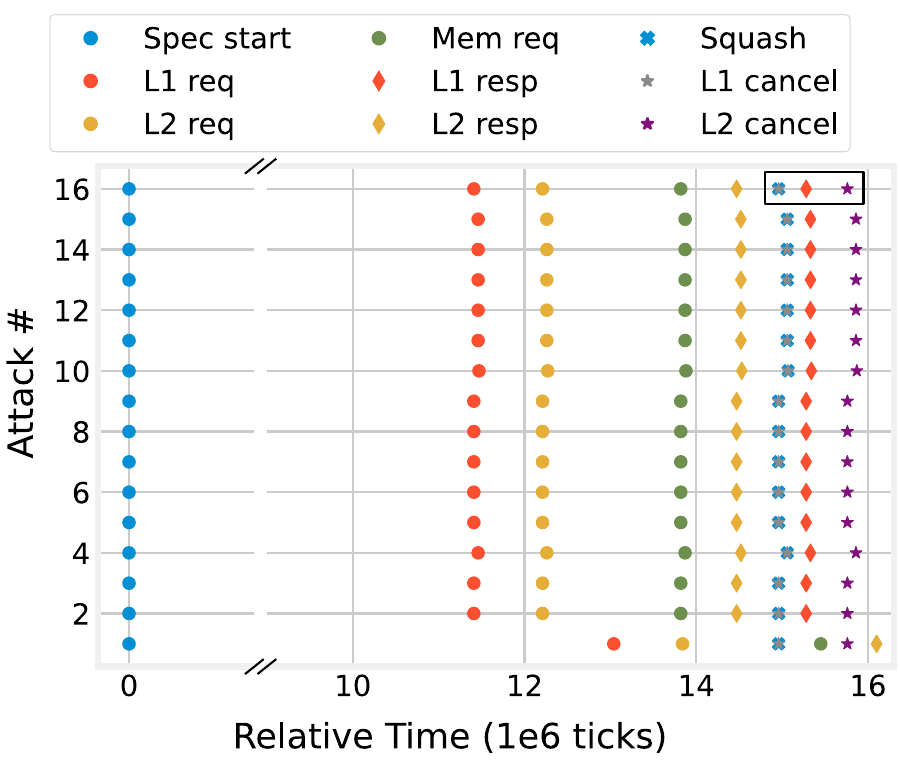}
    \caption{\gls{CPU} events for \textbf{transmit} instructions of all attacks from \textbf{Experiment 4}. In all but the first attack, the cancellation reaches L2 only after the response from memory has arrived.}
    \label{fig:safeside-pht-plot-CMR-forceFail}
\end{figure}

Running the modified Spectre \gls{poc} (with secret caching), we find that the entire secret is indeed extracted. \Cref{fig:safeside-pht-plot-CMR-forceFail} shows the relevant events. The key change, highlighted with a black box, compared to \Cref{fig:safeside-pht-plot-CMR-cachedSecret} is that the response reaches L2 before the cancellation. This is not the case for L1; however, the cache state change in L2 alone is sufficient for the extract step of the Spectre attack to succeed. This is due to the large difference in hit access latencies for L1 and L2 (80 vs. 160, respectively) caused by the large latencies used for C2.

Here, the \gls{poc} requires fewer attacks than in Experiment 1 to extract the secret because we use the modified \gls{poc} where we intentionally cache the secret before each Spectre attack.

\paragraph{Effectiveness.} 
\Cref{tab:cache-changes} shows the cache change metric, $CC$ (\Cref{eq:cache-change}) for all experiments. Experiment 1 gives a value of 1 because \feature{} is not enabled and any transmit instruction executed results in changes to all caches. For experiments 2 and 3, we get a value of 0 because no cache changes occurred. %
Experiment 4 shows that when transmit instructions manage to cause a partial change to cache state, the value of $CC$ is between 0 and 1.
\definecolor{applegreen}{rgb}{0.55, 0.71, 0.0}
\begin{table}[ht]
    \centering %
    \begin{tabular}{@{}lrrrrrr@{}}
    \toprule
        \textbf{Experiment} & \textbf{$N_{1}$} & \textbf{$N_{2}$} & \textbf{$N_{total}$} & \textbf{CC} \\
        \midrule
        1           & \color{red}{29}        & \color{red}{29}        & 29  & \color{red}{1}      \\
        2           & \color{applegreen}{0}  & \color{applegreen}{0}  & 104 & \color{applegreen}{0}      \\
        3           & \color{applegreen}{0}  & \color{applegreen}{0}  & 204 & \color{applegreen}{0}      \\
        4           & 0                      & \color{red}{15}        & 32  & \color{red}{0.234}  \\ \bottomrule
    \end{tabular} %
    \caption{The values of $CC$ for all experiments.}
    \label{tab:cache-changes}
\end{table}

\subsubsection{Google SafeSide -- ret2spec}
As before, we evaluate \feature{} on the \texttt{ret2spec\_sa} \gls{poc} and verify that it thwarts the attack: no part of the secret is leaked. Ret2spec, also called Spectre-RSB~\cite{spectreRSB,transientfail}, targets the \gls{RSB}, which predicts the addresses of return instructions. It exploits the fact that the \gls{RSB} has a limited number of entries, and must remove addresses once it is filled due to deeply nested functions, causing mispredictions of the return addresses. These mispredictions lead to speculatively executed instructions that are abused by ret2spec to leak the secret.

\subsection{Security limitations}

\subsubsection{Cache hits for transmit instructions}
Flush+Reload uses flushing to prepare caches before transmit instructions are executed, preventing the cache line from being present at any level. 
However, an attacker can instead use evictions~\cite{primeProbe} to remove the data from one level but keep it in lower levels. For example, an attacker can evict the data from L1 but not L2 before launching the Spectre attack, and later use the timing difference between an L1 access and an L2 access to leak the secret. While \feature{} is less effective against such attacks,  they are also more difficult to launch due to the smaller timing difference. Further, for architecturally inaccessible secrets that are flushed out on context switches, the attacker must first successfully cache the secret, which is made difficult with \feature{} (see Experiment 2).%

\subsubsection{Windowing gadgets}
We demonstrated the effectiveness of \feature{} against Spectre \glspl{poc}, which were originally created to simply show the feasibility of the attack. Attackers can use several techniques to make attacks more robust. One is to increase the speculation window~\cite{microsoftWindowingGadgets}. Gigerl~\cite{gigerlAutomatedAnalysisSpeculation}, Mambretti et al.~\cite{speculator} and Xiao et al.~\cite{speechminer} identify empirical limits on speculation window sizes achievable on different platforms via different instructions for speculation conditions.

While increasing speculation window size can reduce the effectiveness of \feature{}, attackers must also overcome other practical challenges that maintain \feature{}'s effectiveness. In the \glspl{poc}, both the victim and attacker code are under our control. In practice, attackers are forced to rely on speculative \emph{gadgets}~\cite{intelDisclosure,kasper,inspectreGadget,specrop} present in the victim's code, in a manner similar to \gls{ROP} gadgets~\cite{rop2007}. This includes \emph{disclosure} gadgets, used to access and transmit the secret, and \emph{windowing} gadgets, needed to increase the speculation window as described above. Significant prior work has been done to investigate and reduce the availability of speculative gadgets in critical software targets such as the Linux Kernel~\cite{kasper,intelIBT}. As a result, 
\begin{inparaenum}
    \item attackers are increasingly forced to use less-than-ideal \emph{disclosure} gadgets that can have many redundant instructions (reducing the effective available speculation window), and
    \item fewer windowing gadgets are available, making it harder to circumvent \feature{}.
\end{inparaenum}
\feature{} serves as complementary work to reduce the effective attack surface of critical software targets.

\subsubsection{Speculative-interference attacks}
A crucial requirement for \feature{} is that the transmit instruction is speculative, and is therefore squashed when the speculation window ends. In speculative interference attacks~\cite{specInterference}, however, this is not the case. Instead, speculative execution is used to affect the order of \emph{non-speculative} loads/stores, resulting in a cache state difference that can later be measured to leak the secret. As non-speculative memory requests cannot be cancelled, \feature{} cannot thwart such attacks. We consider them out-of-scope and rely on orthogonal defenses, e.g., full pipelining to prevent speculative instructions from affecting older ones, as suggested by Behnia et al~\cite{specInterference}.

\subsubsection{Non-cache-based side channels}
\feature{} works by reducing \emph{persistent} secret-dependent changes to \emph{cache state}. As a result, \feature{} only covers cache-based side channels. Other side channels, e.g. contention-based channels~\cite{secsmt,binoculars2022,paccagnellaLordRingSide2021}, are out-of-scope, as in many invisible speculation schemes (\Cref{sec:related}).

\section{Performance Evaluation}\label{sec:perf}

\begin{figure*}[t]
\footnotesize
\begin{centering}
    \begin{minipage}[t]{0.33\textwidth}
    \begin{subfigure}{\linewidth}
    \centering
        \noindent\includegraphics[height=5.75cm]{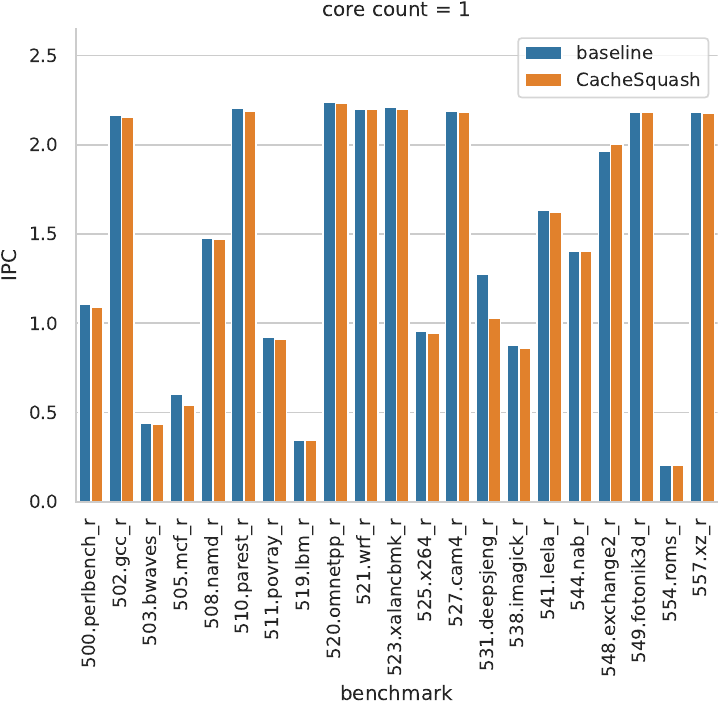}
        \caption{SPEC CPU 2017 1-core}\label{fig:spec1}
    \end{subfigure}
    
    \begin{subfigure}{\linewidth}
    \centering
        \noindent\includegraphics[height=5.75cm]{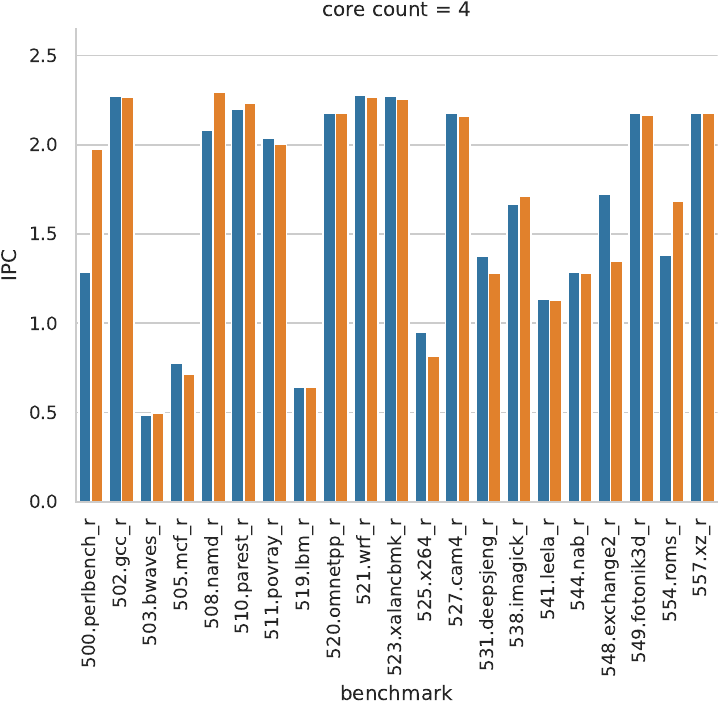}
        \caption{SPEC CPU 2017 4-core}\label{fig:spec4}
    \end{subfigure}
    \end{minipage}%
    \begin{minipage}[t]{0.33\textwidth}
    \begin{subfigure}{\linewidth}
    \centering
        \noindent\includegraphics[height=5.75cm]{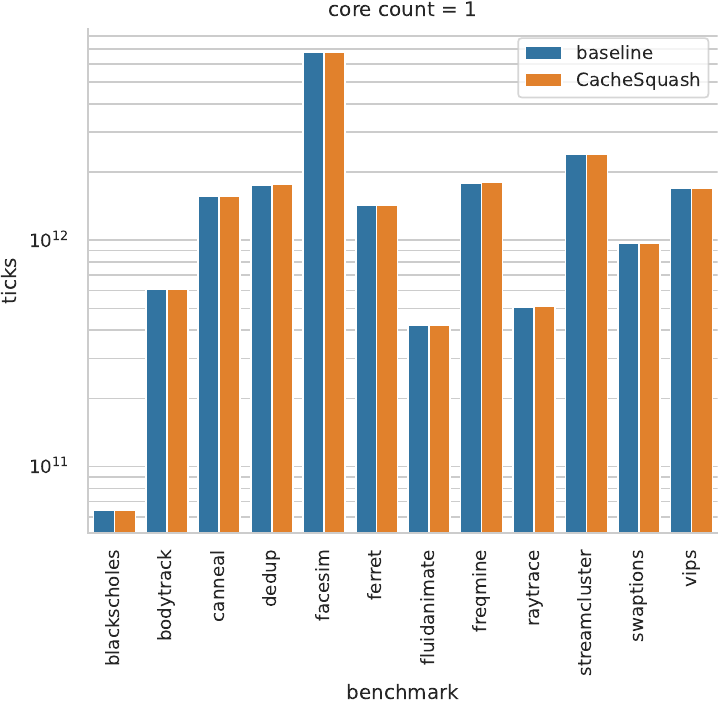}
        \caption{PARSEC 1-core medium input size}\label{fig:parsec-med1}
    \end{subfigure}
    
    \begin{subfigure}{\linewidth}
    \centering
        \noindent\includegraphics[height=5.75cm]{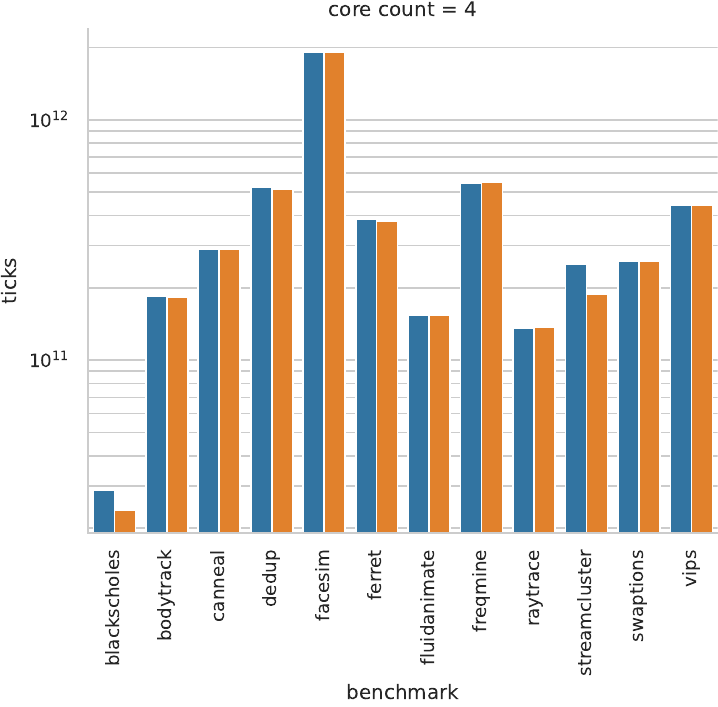}
        \caption{PARSEC 4-core medium input size}\label{fig:parsec-med4}
    \end{subfigure}
    \end{minipage}%
    \begin{minipage}[t]{0.33\textwidth}
    \begin{subfigure}{\linewidth}
    \centering
        \noindent\includegraphics[height=5.75cm]{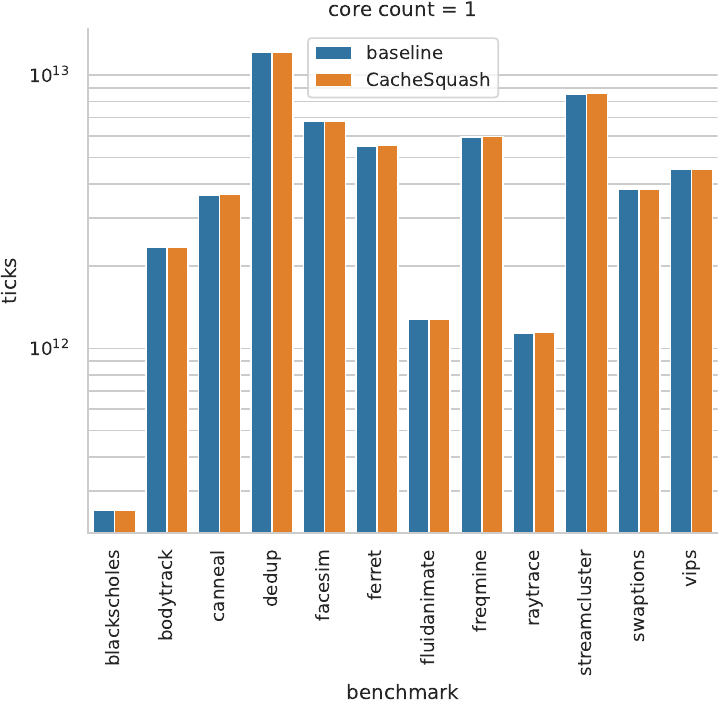}
        \caption{PARSEC 1-core large input size}\label{fig:parsec-large1}
    \end{subfigure}
    
    \begin{subfigure}{\linewidth}
    \centering
        \noindent\includegraphics[height=5.75cm]{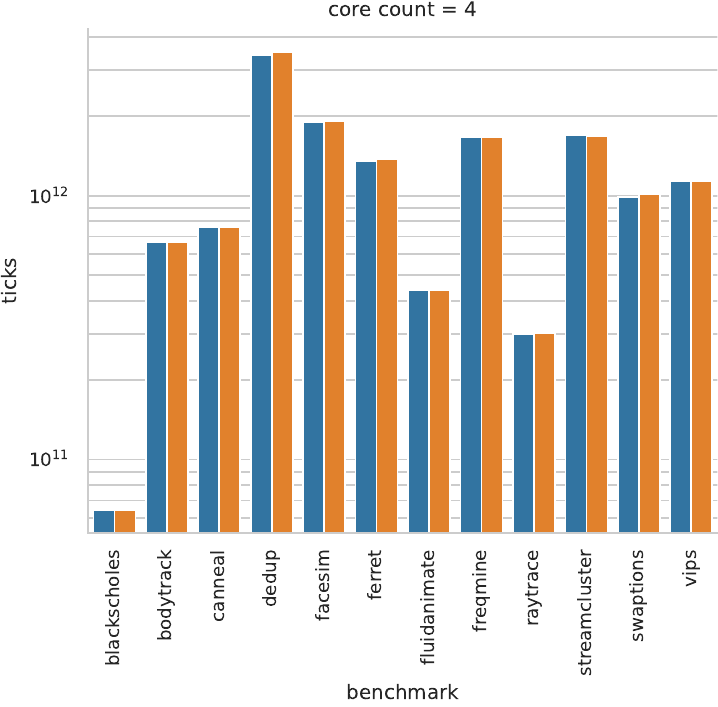}
        \caption{PARSEC 4-core large input size}\label{fig:parsec-large4}
    \end{subfigure}
    \end{minipage}
    \caption{Results for SPEC CPU 2017 (\gls{IPC}) with \texttt{ref} input size and PARSEC (ticks) with medium and large input sizes on 1- and 4-core configurations of baseline and \feature.}
    \label{fig:perf}
\end{centering}
\end{figure*}

We evaluate \feature's performance on 1- and 4-core configurations using similar parameters to Invisispec~\cite{} and CleanupSpec~\cite{cleanupSpec}. We first run the SPEC CPU 2017 benchmarks~\cite{spec17} with the \texttt{ref} input size, using a warm-up period of 10B instructions and measuring the \gls{IPC} for the next 1B instructions. This follows standard procedure from prior work~\cite{STT,invisiSpec,cleanupSpec}. We exclude \texttt{507.cactuBSSN\_r} as it crashes on the baseline and \feature{}. The results are shown in \Cref{fig:spec1,fig:spec4}. The geometric mean \gls{IPC} overhead across all benchmarks and configurations is \GeomeanIpcSpecAllCores. However, the results show a high geometric standard deviation of \GstdSpecAllCores. As our simulations are deterministic, running the benchmarks more than once produces identical results. We therefore evaluate further using the PARSEC benchmarks~\cite{bienia_parsec_2008}. We run them to completion with the medium and large input sizes and measure the number of gem5 ticks (\Cref{fig:parsec-med1,fig:parsec-med4,fig:parsec-large1,fig:parsec-large4}) yielding a geometric mean \emph{speedup} of \GeomeanTicksSpeedupParsecMedAllCores and overhead of \GeomeanTicksOverheadParsecLargeAllCores with lower geometric standard deviations of \GstdParsecMedAllCores and \GstdParsecLargeAllCores, respectively.

\begin{table}[th]
    \centering %
    \begin{tabular}{@{}lr@{}}
    \toprule
        \textbf{Parameter} & \textbf{Value} \\
        \midrule
        Core count                      & 1,4      \\
        Core frequency (GHz)            & 3        \\
        Private  L1I/L1D size (kB)      & 32/64    \\
        Shared L2 size per core (MB)    & 2        \\
        L1/L2 associativity             & 2/8      \\
        L1I/L1D/L2 latency (cycles)     & 1/2/20   \\
        \bottomrule
    \end{tabular} %
    \caption{Parameters used in performance evaluation.}
    \label{tab:perf-configs}
\end{table}

\section{Discussion \& future work}

\paragraph{Meltdown.}
\feature{} provides support for read request cancellations \emph{regardless of the reason for cancellation}. While we tackle speculative execution in this paper, \feature{} is also applicable to fault-based transient attacks such as Meltdown~\cite{Lipp2018meltdown}. Any transmit instruction executed transiently during the Meltdown attack can be cancelled once the fault is detected, thereby reducing cache state changes and reducing the attack's chance of success.

\paragraph{Cancellation broadcasts.} %
Dedicated circuitry, similar to that used for snooping, can be added to the \gls{CPU} die to broadcast cancellations to all caches, even if a snooping protocol is not used. This can drastically improve the security provided by \feature{}, by eliminating the dependency on cache forwarding latency. However, this adds complexity to cancellation handling because lower-level caches would now get cancellations even if the corresponding upstream \gls{MSHR} is not empty. A mechanism must therefore be added to allow lower-level caches to track upstream \glspl{MSHR} and only act on cancellations once the upstream \gls{MSHR} is deallocated.

\paragraph{Cancellation of memory bus transactions.}
In \Cref{sec:goals}, we explicitly avoid changes to external modules and interfaces such as main memory (\hcreq) to enable backward compatibility of \feature{}. However, introducing cancellations to memory buses can be an opportunity to improve system performance. By aborting transactions that are no longer needed, we can free up the memory bus for other transactions. Furthermore, for memories with integrated on-chip caches, this can improve security by cancelling changes to the on-chip cache. We leave such research for future work.

\paragraph{Overlapping cancelled and uncancelled requests.}
If there are $n$ requests waiting for the same cache block, canceling up to $(n-1)$ of them will have no effect on the cache as the MSHR must still be serviced. Information leakage can occur if the \emph{first} request to allocate the \gls{MSHR} is cancelled, but the \gls{MSHR} cannot be deallocated due to the existence of other non-speculative targets that arrived later. There is a timing difference between when the response arrives in this case, and when the response would have arrived had there not been the first request (i.e., the \gls{MSHR} was instead allocated by the non-speculative second request). While this timing difference can theoretically be used to leak information, our threat model assumes that this second request is not under the attacker's control, and they cannot determine the time at which it occurs (and therefore cannot accurately measure the timing difference). Note that if the attacker can control this non-speculative request, they have no need for a Spectre attack and can simply use a non-transient cache timing attack.

\section{Related work}\label{sec:related}

Invisible speculation mechanisms~\cite{invisiSpec,safespec,delay-on-miss,conditionalSpeculation,muontrap} attempt to hide speculative side effects until they are determined to be non-speculative. While hidden, speculative changes are stored in shadow structures which are invisible to the rest of the non-speculative system. CleanupSpec~\cite{cleanupSpec} takes an ``undo'', rather than a hiding, approach, allowing speculative changes to be seen by the system, but undoing them on squashes. However, prior schemes suffer from significant overheads, (e.g., $21-72\%$ for InvisiSpec) and/or require the addition of expensive on-chip storage to track speculative changes (e.g., L0 in MuonTrap, buffers in CleanupSpec).  In comparison, \feature{} does not require any structures to track speculative changes and has negligible overheads.

\Gls{STT}~\cite{STT} is another Spectre mitigation technique that taints data loaded by Spectre access instructions and delays any instructions that use it until the access instruction becomes non-speculative. While \gls{STT} is not limited to protecting only cache-based side channels, it can result in significant overheads ($8.5-14.5\%$) compared to \feature{} and does not cover the case where only the transmit, but not the access, instruction is speculative.

\ifAnon
\else
\section*{Acknowledgments}
This work is supported in part by Natural Sciences and Engineering Research Council of Canada (grant number RGPIN-2020-04744). Views expressed in the paper are those of the authors and do not necessarily reflect the position of the funders.
\fi

\bibliographystyle{IEEEtran}
\bibliography{refs}

\end{document}